# Assessing and Comparing the Coverage of Italian Publications in OpenCitations: a Study within Six Italian Universities

Erica Andreose[1] [orcid:0009-0003-7124-9639], Ivan Heibi[2,3] [orcid:0000-0001-5366-5194], Silvio Peroni[2,3] [orcid:0000-0003-0530-4305], Leonardo Zilli[1] [orcid:0009-0007-4127-4875]

[1] Digital Humanities and Digital Knowledge, Department of Classical Philology and Italian Studies, University of Bologna, Bologna, Italy

[2] Research Centre for Open Scholarly Metadata, Department of Classical Philology and Italian Studies, University of Bologna, Bologna, Italy

[3] Digital Humanities Advanced Research Centre (/DH.arc), Department of Classical Philology and Italian Studies, University of Bologna, Bologna, Italy

## Abstract

Recent initiatives advocating responsible, transparent research assessment have intensified the call to use open research information rather than proprietary databases. This study evaluates the coverage and citation representation of publications recorded in the Current Research Information Systems (CRIS), all instances of the IRIS software platform, of six Italian universities within OpenCitations, a community-owned open infrastructure. Using persistent identifiers (DOIs, PMIDs, and ISBNs) specified in the IRIS installations involved, we matched the publications recorded in OpenCitations Meta and extracted the related citation links from the OpenCitations Index. Results show that OpenCitations covers, on average, over 40% of IRIS publications, which is quantitatively comparable to those reported by Scopus and Web of Science in another study. However, gaps persist, particularly for publication types prevalent in the Social Sciences and Humanities, such as monographs and critical editions. Overall, the findings demonstrate the growing maturity of OpenCitations and, more broadly, of Open Science infrastructures as viable alternatives as sources of research information, while highlighting areas where further metadata enrichment and interoperability efforts are needed.

## Introduction

A growing body of research has examined the tensions between open and commercial data ecosystems, highlighting differences in access, governance, and value creation (Janssen, Charalabidis, & Zuiderwijk, 2012). Notable previous works have emphasised the open vs proprietary data usage in several fields, discussing the challenges of data sharing and reuse, the development of open scholarly infrastructures, and the continued dominance of proprietary bibliometric systems in research assessment (Borgman, 2015; Larivière, Haustein, & Mongeon, 2015; Tennant et al., 2019).

The San Francisco Declaration on Research Assessment (2013) (DORA, https://sfdora.org), the Leiden Manifesto (https://www.leidenmanifesto.org/) (Hicks et al., 2025), the Coalition for Advancing Research Assessment (2022) (CoARA, https://coara.eu), the Barcelona Declaration on Open Research Information (2024a) (DORI, https://barcelona-declaration.org), are just a few of the recent initiatives that have pushed for having more transparent research assessment procedures. Such transparency is reachable, of course, by publishing the methodologies adopted for the assessment that must, however, be based on *open research information* (i.e. metadata relating to the conduct and communication of research, such as bibliographic metadata, citation data, information on funding and grants, and information on organisations and research contributors.

Indeed, to address the latter aspect, DORI members have started organising joint work to identify crucial aspects of the implementation and adoption of open research information sources within research-performing organisations. These joint works have resulted in the creation of several working groups (WGs), each originating from one point of the roadmap (https://barcelona-declaration.org/roadmap/) listing the primary properties identified by DORI members (Barcelona Declaration on Open Research Information, 2024b) that include:

- [WG1] journal article metadata and book metadata;
- [WG2] metadata of research outputs in repositories;
- [WG3] funding metadata;
- [WG4] replacing closed systems;
- [WG5] sustaining infrastructures;
- [WG6] evaluating sources of open research information;
- [WG7] evidence of benefits.

The study presented in this article spans several activities related to these WGs, in particular WGs 4 and 6. Indeed, recent literature has shown that open scholarly infrastructures and other open research information sources have begun to exert a significant influence in several studies within and beyond the field of quantitative science studies (Cao et al., 2026). However, there are still barriers to the adoption of open research information at large, mainly derived by the lack of institutional resources and streamlined processes to facilitate its use (Danesh & Rahimi, 2026), together with other issues related to the current use of global rankings based on proprietary sources, the still minimal number of implemented policies to regulate the use of open research information at institutional level, and, more generally, a perceived mistrust of open data. As recently reported by Kuchma et al. (2025), in the context of the CoARA Working Group *Towards Open Infrastructure for Responsible Research Assessment* (OI4RRA, https://www.coara.org/working-groups/wg-towards-open-infrastructures-for-responsible-research-assessment-oi4rra/), several countries still strongly depend, for research assessment exercises, on data in established commercial databases such as Web of Science (WoS) and Scopus, held by Clarivate Analytics and Elsevier. However, recently, some institutions have indeed started to push for the adoption of open research information sources by unsubscribing from contracts with well-known proprietary sources – such as in the case of the Sorbonne University, which unsubscribed from Web of Science (https://www.sorbonne-universite.fr/en/news/sorbonne-university-unsubscribes-web-science) and the CNRS, which unsubscribed from Scopus (https://www.cnrs.fr/en/update/cnrs-has-unsubscribed-scopus-publications-database).

Within this context, some past studies provide evidence of the appropriateness of the available open research information sources, particularly when compared with closed, proprietary providers. Indeed, replacing closed systems with open sources needs to pass through necessary checks to assess their quality, coverage, openness, and transparency.

For instance, Bologna et al. (2022) examine whether open bibliographic data can substitute for proprietary databases in research assessment, finding that open research information sources are already suitable for some scientific disciplines, even if they remain inadequate for many fields within the humanities.

Indeed, there are other running investigations, for instance, that are guided by DORI WGs, which set up actions to assess if existing open data sources can be comparable in coverage and data quality with the proprietary sources currently adopted, to provide actual clues that a switch from proprietary to open is, indeed, already possible. In this framework, we have recently analysed the Current Research Information System (CRIS) of the University of Bologna (UNIBO), and shown that, from a purely quantitative perspective, there was a comparable coverage in the number of bibliographic resources described in the UNIBO CRIS included across the systems analysed – OpenCitations (open), Scopus (proprietary), and Web of Science (proprietary) – as well as in the number of citations they receive (Andreose et al., 2026a).

In this article, we want to run another analysis on the Italian academic system by considering a few Current Research Information Systems (CRIS) used by Italian universities, which are based on a specific software (i.e. IRIS) adopted at the national level, and the data available in a well-known open scholarly infrastructure (i.e. OpenCitations). To broaden the discussion above, some Italian universities met in Milan in July 2025 to explore potential strategies for further investigating the coverage of their research information in existing open infrastructures. As a consequence of that meeting, we have proposed to run an exploratory study to investigate the following research questions (RQs):

1. What is the coverage in OpenCitations of the publications listed in the IRIS installations of Italian Universities?
2. How many citations are available in OpenCitations that involve, either as the citing entity or as the cited entity, the publications in such IRIS installations?
3. Which types of publications are not covered in OpenCitations?

To answer these questions, we have developed a methodology that builds on the approach we adopted in a previous study (Andreose et al., 2026a) and implemented it in a Python library to ensure experimental repeatability (Zilli et al., 2025). In addition, all data produced by our analysis are available online on Zenodo (Andreose et al., 2026b).

The rest of the article is organised as follows. In Section "Methodology", we introduce the data used and the workflow adopted for the analysis. In Section "Results", we address the RQs by supporting our findings with appropriate data and figures resulting from the analysis. In Section "Discussions", we discuss the results of the study and compare them with prior work on related topics. Finally, in Section "Conclusions", we conclude the article by sketching out some future developments.

# Methodology

In this section, we introduce the methodological steps we followed in the study. In particular, we introduce the data we have collected to support our work, then detail the workflow we devised for the analysis and the software we created to automate it.

**Open Research Information providers adopted**

As anticipated from the introduction, all the data gathered come from two distinct providers of open research information: installations of IRIS in six distinct universities and OpenCitations. IRIS (Bollini et al., 2016) is a software for implementing CRIS instances developed by CINECA, a consortium of Italian universities and research institutions. It is widely adopted by most Italian universities to manage institutional research information, enabling the collection and curation of bibliographic metadata describing scholarly output (e.g. titles, authors, publication venues, and persistent identifiers). IRIS also plays a key role in supporting research assessment exercises, both at a local level, i.e. within a specific university, and at a national level, since it can interact with the system used by the National Agency for the Evaluation of Universities and Research (in Italian: *Agenzia Nazionale per la Valutazione dell'Università e della Ricerca*, ANVUR, https://www.anvur.it/en). Participation in these assessments, as well as access to research funding programs, requires institutions to collect data on their research outputs, contributing to IRIS achieving a high level of coverage of institutional research outputs (Galimberti & Mornati, 2017). In this context, IRIS implementations across institutions adopt shared data models but may include local extensions. In particular, to enable interoperability, descriptive types for publications in IRIS installations are also aligned with the national publication type classification taxonomy of the Ministry of University and Research (MIUR), which provides a standard taxonomy used for research assessment and reporting across Italian institutions.

OpenCitations (https://opencitations.net) (Peroni & Shotton, 2020) is a community-governed open scholarly infrastructure that provides free access to global bibliographic and citation data. Its main collections include OpenCitations Meta (Massari et al., 2024), which stores bibliographic metadata for scholarly resources, and the OpenCitations Index (Heibi et al., 2024), which collects more than 2.4 billion citation links (as of 13 April 2026), which are described as first-class data entities with their own metadata specified. The infrastructure integrates data from multiple sources (as of 13 April 2026: Crossref, DataCite, the National Institute of Health - Open Citation Collection, OpenAIRE, and the Japan Link Center) and releases them under a CC0 licence, enabling unrestricted reuse and interoperability via Semantic Web technologies (Daquino et al., 2020).

**Gathering the data**

Using the mailing list created for the workshop held in Milan in July 2025, mentioned in Section "Introduction", we have asked the participating institutions to be included in our analysis of Italian universities. The request was to provide a subset of data extracted from their IRIS installations, generated by 7 specific SQL queries (described in Appendix 1 for reproducibility). While provided as file dumps for our convenience, all data can be retrieved directly from the related IRIS installations available online. In addition, we have required a mapping document that links the publication types used in each IRIS installation to their corresponding nationwide taxonomy proposed by the MIUR. This document is necessary to align results and enable comparison across all participating universities.

The results of the SQL queries run against each IRIS installation were stored in seven distinct files containing, respectively, the following information:

- data (internal ID, ORCID, name, etc.) about each actor (authors, editors, etc.) involved in the IRIS dataset;
- the list of authors and author count for each record;
- identifiers of each publication, such as DOIs, PMIDs, ISBNs and others;
- language of the publication (if applicable);
- basic bibliographic metadata (title and publication date);
- the names and locations of the publishers of each publication;
- additional metadata regarding the publication context (venue, editors, etc.).

Six different institutions positively agreed to be part of this analysis, i.e. the University of Bologna (UNIBO), the University of Milan (UNIMI)[1], the University of Turin (UNITO), the University of Padua (UNIPD), the University of Eastern Piedmont (UPO), and the Scuola Normale Superiore (SNS) in Pisa. While these six universities are primarily located in the Northern part of Italy, and thus there is no representation of universities in the Central and Southern parts of the country, they are adeguately distributed in terms of their research personnel (i.e. professors, tenure and non-tenure researchers, including research fellows, and adjunct professors), according to the national statistics dated 2024 made available by the MIUR at https://ustat.mur.gov.it/dati/didattica/italia/atenei#tabistituti (consulted on 10 April 2026). In particular, UNIBO and UNIPD have more than 5,000 researchers (6,174 and 5,923, respectively), UNIMI and UNITO between 2,000 and 5,000 researchers (4,653 and 4,223, respectively), and UPO and SNS have fewer than 2,000 researchers (773 and 273, respectively), thus having a balanced representation of three different sizes of Italian universities and sufficiently representative for our study.

All source files obtained from the participating universities were first converted to CSV format to facilitate processing across the IRIS exports. In addition, all publication records were aligned with the MIUR publication-type classification, which serves as a standard reference framework across Italian CRIS systems, using the institutions' provided mappings. Where additional publication types were defined in specific IRIS instantiations (*PhD thesis*, *bachelor thesis*, *master thesis*, etc.) that had no correspondence to the MIUR publication-type classification, we mapped them to the residual category *other*. Table 1 shows the distribution of the top 10 publication types across the six participating institutions.

**Table 1.** Distribution of the top 10 publication types across the participating universities. Percentages represent the share of IRIS records associated with each MIUR publication type relative to the total number of records for each university. The “Other (MIUR)” category is the residual category we used when either an IRIS installation specified a generic “other”

[1] For the specific case of UNIMI, we needed to request additional information from their IRIS installation because of their internal hierarchical category structure for publication types.

publication type as a residual category, or no correspondence was found between a particular IRIS publication type and the MIUR publication-type classification.

| | UNIBO | UNIPD | UNITO | UNIMI | UPO | SNS |
|---|---|---|---|---|---|---|
| *Article in academic journal* | 54.2% | 56.3% | 50.0% | 60.0% | 60.5% | 60.8% |
| *Contribution to volume, chapters or articles* | 14.1% | 11.7% | 15.0% | 10.7% | 13.4% | 12.7% |
| *Contribution in proceedings* | 10.9% | 15.1% | 11.0% | 3.7% | 8.4% | 6.4% |
| *Other (MIUR)* | 2.3% | 4.2% | 3.9% | 11.2% | 2.8% | 4.7% |
| *Monograph or scientific treatise* | 2.5% | 2.4% | 2.9% | 1.9% | 2.4% | 3.1% |
| *Abstract in proceedings* | 5.3% | 3.1% | 2.6% | 1.0% | 2.4% | 0.7% |
| *Review in journal* | 1.9% | 0.9% | 2.5% | 4.1% | 2.1% | 2.5% |
| *Editorship* | 2.0% | 1.1% | 2.0% | 1.5% | 1.7% | 1.8% |
| *Entries in prestigious encyclopedias* | 1.0% | 0.5% | 1.5% | 0.7% | 0.9% | 2.5% |
| *Abstract in journal* | 1.2% | 1.9% | 0.5% | 0.1% | 1.4% | 0.2% |

In addition to the data provided by the institutions, we have reused all data from OpenCitations. In particular, we retrieved the bibliographic metadata for publication entities from OpenCitations Meta (Massari et al., 2024); we used the June 2025 CSV dump (OpenCitations, 2025b), which contains more than 124 million bibliographic entities and more than 1 million publication venues. Also, we retrieved citation data for publication entities from the OpenCitations Index (Heibi et al., 2024), using the July 2025 CSV dump (OpenCitations, 2025a), which contains more than 2.2 billion citation links.

**Data analysis workflow**

For each IRIS data dump, we joined the ITEM_MASTER_ALL table (containing the publications' basic bibliographic metadata, obtained running the Query 5 in Appendix 1) and ITEM_IDENTIFIER table (containing the publications' identifiers, obtained running the Query 3 in Appendix 1) using the internal IRIS identifier field (*ITEM_ID*) to obtain a single core dataset of records, briefly described in Table 2. From this joined table, we extracted all records with *valid* (i.e., syntactically correct) persistent identifiers (PIDs) compliant with one of the following PID schemes, all of which are handled in OpenCitations Meta: DOI, PMID, or ISBN. All records with invalid identifiers (non-numeric, null, or not compliant with the aforementioned PID schemes) were discarded.

**Table 2.** Structure of the core IRIS dataset used in the mapping process, obtained by joining ITEM_MASTER_ALL and ITEM_IDENTIFIER tables from each IRIS dump. For each field, the source table, a brief description, and an illustrative example are provided.

| Source table | Field | Description | Example |
|---|---|---|---|
| ITEM_MASTER_ALL | ITEM_ID | Unique internal identifier assigned to each record in the IRIS installation | 164257 |
| | DATE_ISSUED_YEAR | Year of publication as recorded in IRIS | 2012 |
| | TITLE | Title of the publication | FaBiO and CiTO: Ontologies for describing bibliographic resources and citations |
| | OWNING_COLLECTION | Numeric code identifying the IRIS publication type | 35 |
| | OWNING_COLLECTION_DES | Human-readable label of the IRIS publication type | 1.01 Articolo in rivista |
| | MIUR_TYPE_CODE | MIUR publication type label assigned to the record | 262 |
| | MIUR_TYPE_NAME | Human-readable label of the MIUR publication type | Articolo in rivista |
| ITEM_IDENTIFIER | IDE_DOI | DOI of the publication, if available | 10.1016/j.websem.2012.08.001 |
| | IDE_ISBN | ISBN of the publication, if available | 8883125150 |
| | IDE_PMID | PubMed identifier of the publication, if available | 22995286 |

To further prevent false-positive matches, ISBNs underwent an additional validation step based on the publication type of the corresponding IRIS record. ISBN identifiers were considered valid only for records associated with a predefined subset of MIUR publication types deemed compatible with ISBN assignment, i.e. *monograph or scientific treatise*, *concordance*, *critical edition*, *publication of unpublished sources*, *scientific commentary*, *book translation*, and *editorship*. This restriction is significant because ISBNs typically identify container entities (e.g. monographs or edited volumes). However, individual chapters or contributions may be submitted to an IRIS instance using the ISBN of their container entity, resulting in these constituent parts often being represented by the container entity's ISBN (in particular, when other identifiers are not specified).

From the validated core IRIS dataset, we extracted a list of all unique PIDs. We then processed the OpenCitations Meta dump to retrieve all corresponding OpenCitations Meta Identifiers (OMIDs) by parsing the *id* field of each OpenCitations Meta record, which encodes all available identifiers as a space-separated string (e.g., `"omid:br/06250314836 doi:10.1177/0971721819841995 openalex:W2944531193"`) to match the PIDs and extract the relevant OMIDs. Then, we extracted from the OpenCitations Index dump all citations that involve any of the OMIDs retrieved.

Finally, we applied a deduplication step to address records that were duplicated upstream in OpenCitations Meta. Because of these duplicates, the same PID can appear in OpenCitations Meta linked to multiple OMIDs, resulting in multiple matched entities for a single IRIS record. Records sharing the same IRIS internal identifier were deduplicated by selecting the entry with the most complete publication date, defined as the date with the highest level of granularity (i.e. *year-month* preferred over *year-only*). When this was insufficient to deduplicate to a single entity (e.g. when multiple duplicates shared the same level of date granularity), the remaining candidates were ordered by OMID in descending order, and the first entry in the ordered sequence was selected.

This deduplication step was intentionally applied after querying the OpenCitations Index, because it contains citation records referencing every entity present in OpenCitations Meta. Using the pre-deduplicated set of OMIDs ensures that the matching process also accounts for duplicated OpenCitations Meta entities, preserving complete coverage during the citation mapping. Finally, all the citations have been deduplicated using their Open Citation Identifier (OCI) (Peroni & Shotton, 2019) and the PIDs specified in each IRIS record in case multiple OMIDs have been assigned to the same publication entity.

**Software package**

To make it easier to run the analysis, we have developed a software tool, i.e. *iris-oc-mapper* (Zilli et al., 2025) (https://github.com/opencitations/iris-oc-mapper), which provides a command-line tool to search bibliographic entities from an IRIS (Institutional Research Information System) dump within OpenCitations Meta and Index data dumps, thus implementing the workflow depicted in the previous subsection. In particular, it allows to:

- convert IRIS dumps into structured and manageable CSV archives;
- map IRIS records types to the types defined by MIUR;
- analyse IRIS dumps to extract relevant bibliographic information;
- map the coverage of IRIS dumps within the OpenCitations Meta and OpenCitations Index collections;
- create sub-datasets of IRIS dumps based on their mapping status (found in OpenCitations Meta, not found in OpenCitations Meta, found in OpenCitations Index, records without persistent identifiers);
- generate HTML reports summarising the analysis and mapping results.

In addition to supporting the work described in this article, the idea of developing such software is to enable the Italian universities to re-run the same process on their data autonomously in the future. Also, we have taken particular care to develop visual HTML-based interfaces that enable domain experts (e.g., university librarians and decision-makers) to grasp the main outcomes of our analysis easily.

The generated HTML report is organised into functional sections that correspond to the different stages of the analysis workflow. An initial overview of the IRIS data ingestion and PID identification is provided, including a Sankey diagram and statistics describing the extraction and validation flow for PIDs (Figure 1).

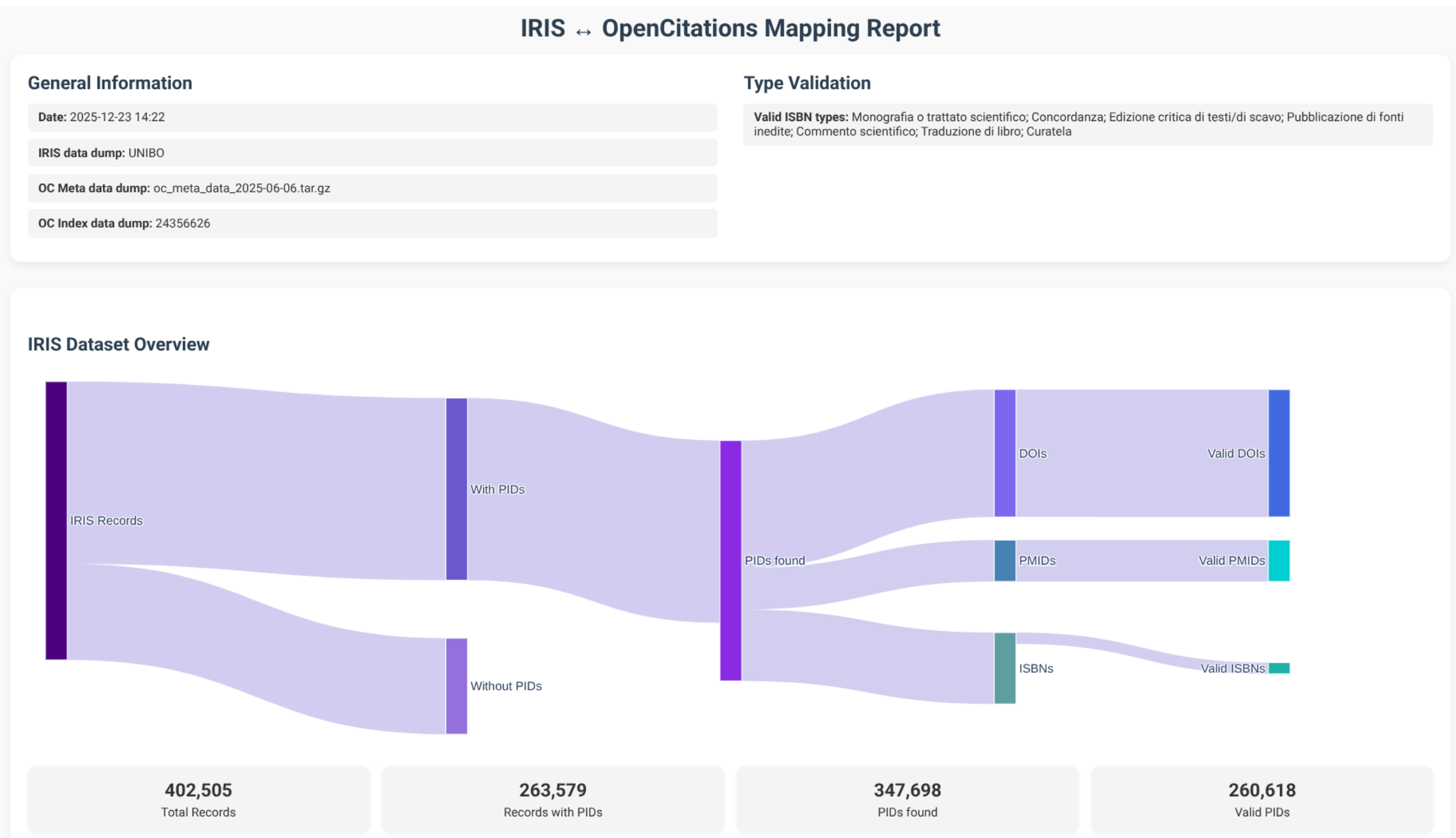

**Figure 1.** First section of one of the HTML reports created by the *iris-oc-mapper* software, displaying mapping run metadata alongside a Sankey diagram and statistics describing the extraction and validation flow of PIDs.

The second section of the HTML report describes records excluded from subsequent processing stages due to missing or malformed metadata. It includes a breakdown of publication types for excluded records and an analysis of invalid or misassigned PIDs by identifier type (Figure 2). A chronological distribution of the IRIS data dump is shown in another diagram, and plots the number of records by publication year (Figure 3).

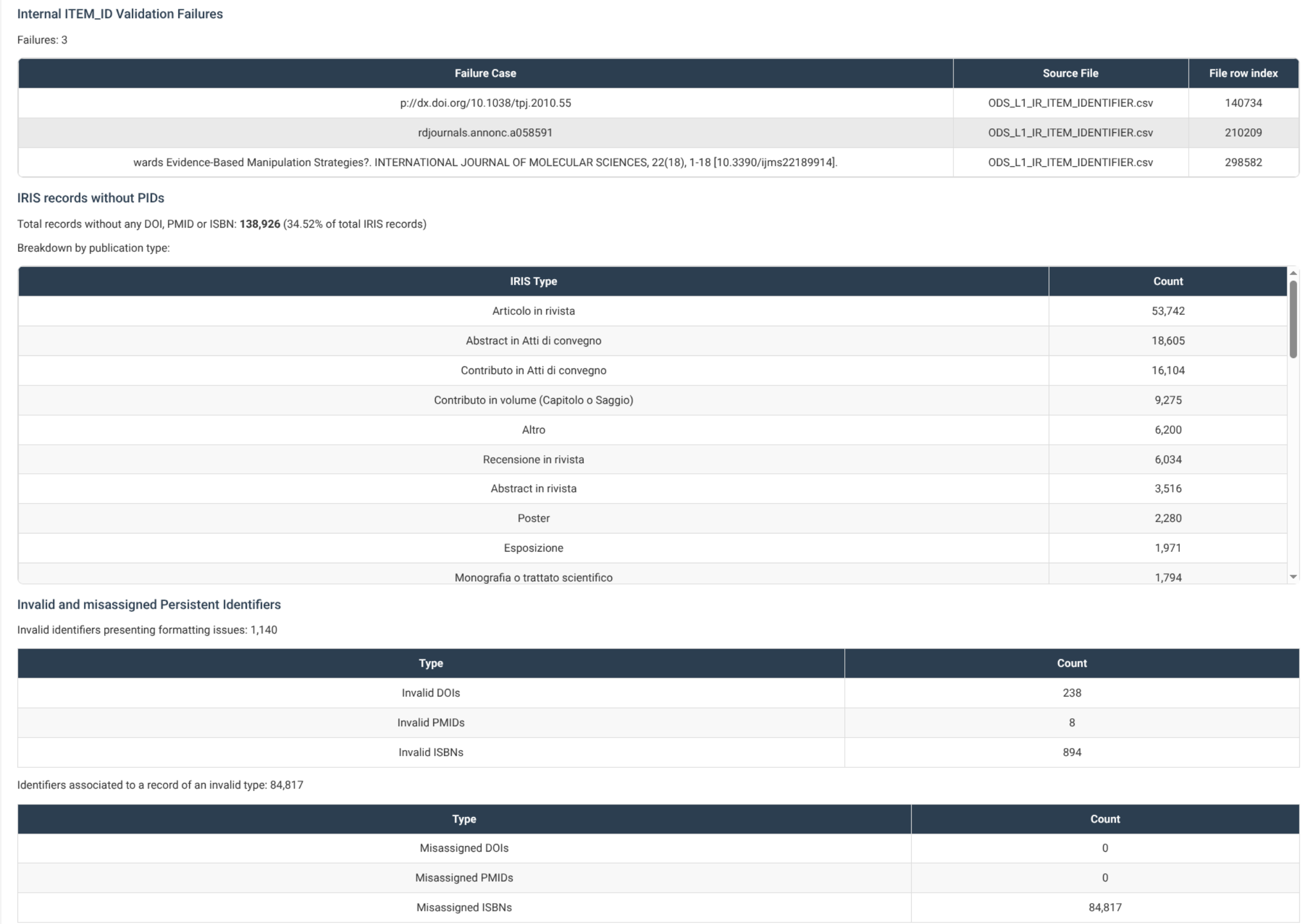

### Internal ITEM_ID Validation Failures

Failures: 3

| Failure Case | Source File | File row index |
|---|---|---|
| p://dx.doi.org/10.1038/tpj.2010.55 | ODS_L1_IR_ITEM_IDENTIFIER.csv | 140734 |
| rdjournals.annonc.a058591 | ODS_L1_IR_ITEM_IDENTIFIER.csv | 210209 |
| wards Evidence-Based Manipulation Strategies?. INTERNATIONAL JOURNAL OF MOLECULAR SCIENCES, 22(18), 1-18 [10.3390/ijms22189914]. | ODS_L1_IR_ITEM_IDENTIFIER.csv | 298582 |

### IRIS records without PIDs

Total records without any DOI, PMID or ISBN: **138,926** (34.52% of total IRIS records)

Breakdown by publication type:

| IRIS Type | Count |
|---|---|
| Articolo in rivista | 53,742 |
| Abstract in Atti di convegno | 18,605 |
| Contributo in Atti di convegno | 16,104 |
| Contributo in volume (Capitolo o Saggio) | 9,275 |
| Altro | 6,200 |
| Recensione in rivista | 6,034 |
| Abstract in rivista | 3,516 |
| Poster | 2,280 |
| Esposizione | 1,971 |
| Monografia o trattato scientifico | 1,794 |

### Invalid and misassigned Persistent Identifiers

Invalid identifiers presenting formatting issues: 1,140

| Type | Count |
|---|---|
| Invalid DOIs | 238 |
| Invalid PMIDs | 8 |
| Invalid ISBNs | 894 |

Identifiers associated to a record of an invalid type: 84,817

| Type | Count |
|---|---|
| Misassigned DOIs | 0 |
| Misassigned PMIDs | 0 |
| Misassigned ISBNs | 84,817 |

**Figure 2.** The second section of the HTML report describes records excluded from subsequent processing stages due to missing or malformed metadata. It includes a breakdown of publication types of excluded records, as well as an analysis of invalid or misassigned PIDs by identifier type.

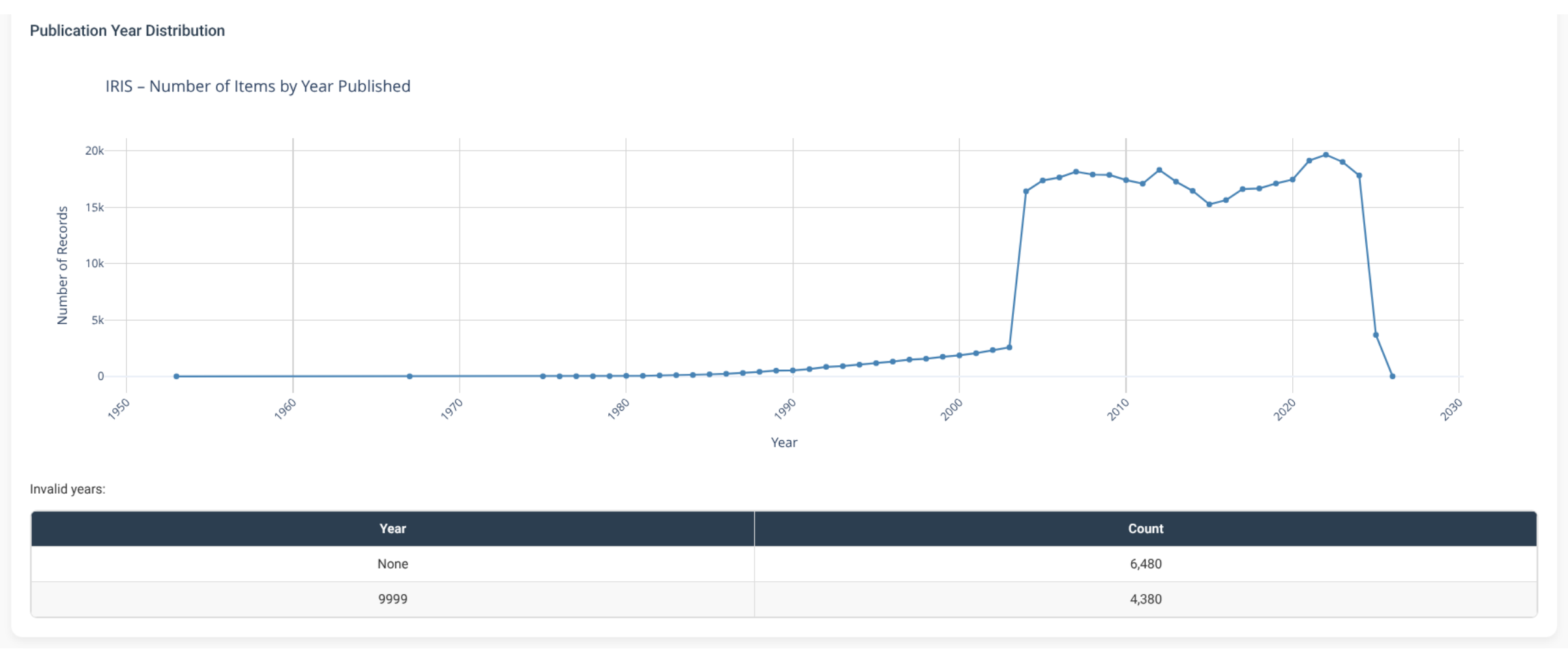

Invalid years:

| Year | Count |
|---|---|
| None | 6,480 |
| 9999 | 4,380 |

**Figure 3.** Temporal visualisation of the IRIS data dump, plotting the number of records by publication year. Invalid year values are also mentioned in a separate table.

Finally, the mapping results quantifying coverage within OpenCitations Meta are presented, providing breakdowns of publication types for both subsets of matched and unmatched records (Figure 4). Finally, the last section of the HTML report provides a summary of the citation analysis conducted using the OpenCitations Index (Figure 5).

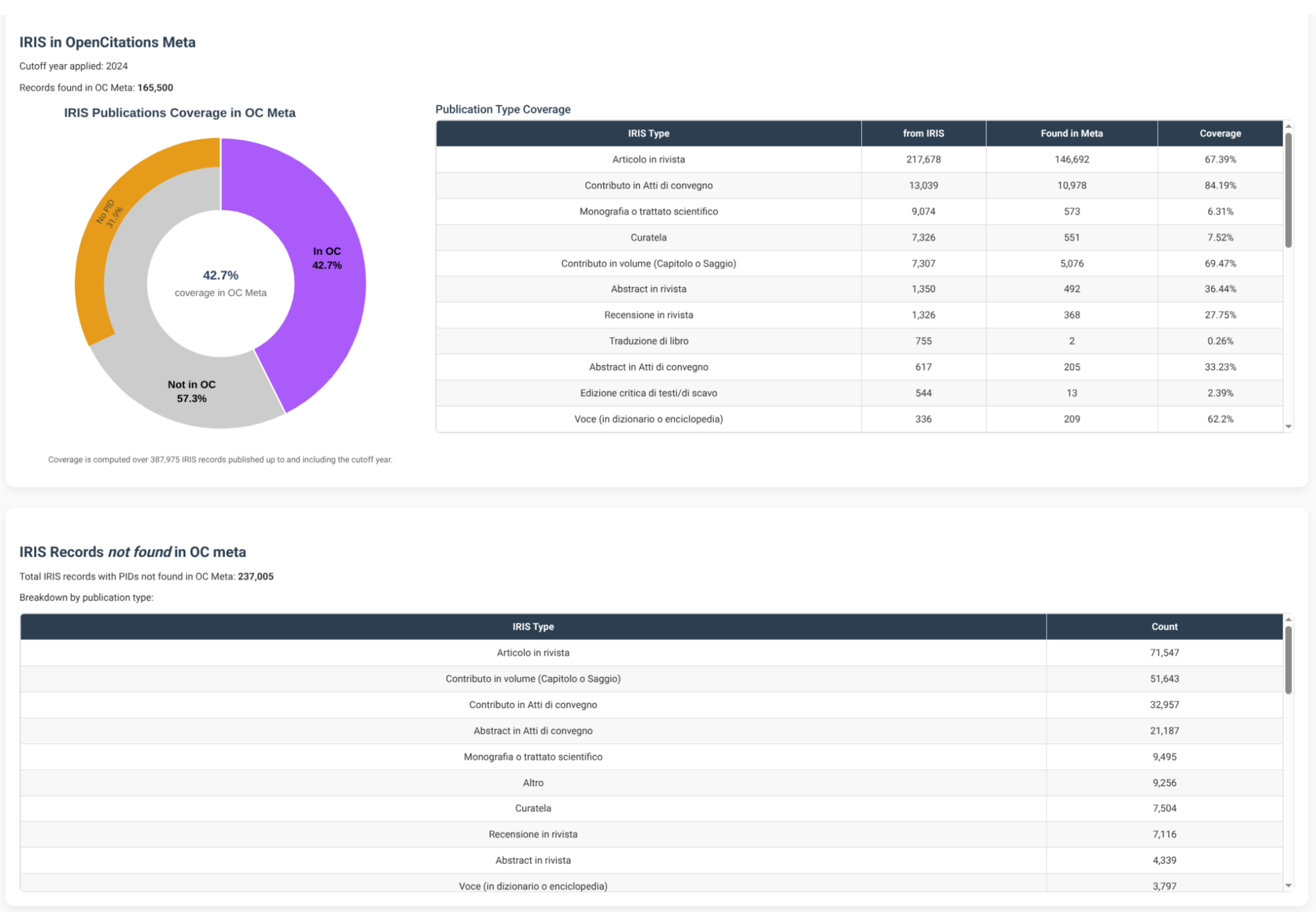

### IRIS in OpenCitations Meta

Cutoff year applied: 2024

Records found in OC Meta: **165,500**

Publication Type Coverage

| IRIS Type | from IRIS | Found in Meta | Coverage |
|---|---|---|---|
| Articolo in rivista | 217,678 | 146,692 | 67.39% |
| Contributo in Atti di convegno | 13,039 | 10,978 | 84.19% |
| Monografia o trattato scientifico | 9,074 | 573 | 6.31% |
| Curatela | 7,326 | 551 | 7.52% |
| Contributo in volume (Capitolo o Saggio) | 7,307 | 5,076 | 69.47% |
| Abstract in rivista | 1,350 | 492 | 36.44% |
| Recensione in rivista | 1,326 | 368 | 27.75% |
| Traduzione di libro | 755 | 2 | 0.26% |
| Abstract in Atti di convegno | 617 | 205 | 33.23% |
| Edizione critica di testi/di scavo | 544 | 13 | 2.39% |
| Voce (in dizionario o enciclopedia) | 336 | 209 | 62.2% |

### IRIS Records *not found* in OC meta

Total IRIS records with PIDs not found in OC Meta: **237,005**

Breakdown by publication type:

| IRIS Type | Count |
|---|---|
| Articolo in rivista | 71,547 |
| Contributo in volume (Capitolo o Saggio) | 51,643 |
| Contributo in Atti di convegno | 32,957 |
| Abstract in Atti di convegno | 21,187 |
| Monografia o trattato scientifico | 9,495 |
| Altro | 9,256 |
| Curatela | 7,504 |
| Recensione in rivista | 7,116 |
| Abstract in rivista | 4,339 |
| Voce (in dizionario o enciclopedia) | 3,797 |

**Figure 4.** The fourth section of the HTML report describes the coverage analysis of IRIS records within OpenCitations Meta. Breakdowns of publication types are provided for both the matched and unmatched record subsets.

### IRIS in OpenCitations Index

Cutoff year applied: 2024

Total number of citations involving IRIS publications: **9,992,645**

| Role | Citation count |
|---|---|
| IRIS records as **citing** entity | 5,263,041 |
| IRIS records as **cited** entity | 5,192,299 |
| Citations **between IRIS** publications | 462,695 |

This report was generated by IRIS-OC Mapper, version 1.0.5.

**Figure 5. The c**oncluding section of the HTML report summarises the results of the citation analysis of matched records.

**Table 3.** Summary of PID extraction and validation across IRIS instances. The field *Total records* indicates the number of records in each IRIS data dump. *Records with PIDs* indicates the number of records having at least one DOI, PMID, or ISBN. *Total PIDs extracted* refers to the total number of raw PIDs found in IRIS. *PIDs by type* reports the distribution of these identifiers across PID categories. *Valid PIDs* counts identifiers retained after validation and deduplication within each IRIS record. *Misassigned ISBNs* indicates the number of ISBNs excluded due to incompatibility with the associated publication type. *Final PID list size* reports the number of identifiers retained for searching within OpenCitations Meta.

| | **UNIBO** | | | **UNIPD** | | | **UNITO** | | | **UNIMI** | | | **UPO** | | | **SNS** | | |
|---|---|---|---|---|---|---|---|---|---|---|---|---|---|---|---|---|---|---|
| *Total records* | 402,505 | | | 416,547 | | | 326,137 | | | 381,525 | | | 62,638 | | | 39,979 | | |
| *Records with PIDs* | 263,579 | | | 230,467 | | | 177,265 | | | 256,454 | | | 35,880 | | | 25,214 | | |
| *Total PIDs extracted* | 347,698 | | | 310,217 | | | 230,304 | | | 378,198 | | | 47,607 | | | 28,388 | | |
| *PID by type (DOI \| PMID \| ISBN)* | 184,466 | 59,984 | 103,248 | 169,790 | 69,353 | 71,074 | 100,073 | 61,067 | 69,164 | 192,543 | 117,237 | 68,418 | 24,632 | 12,083 | 10,892 | 18,023 | 1,302 | 9,063 |
| *Valid PIDs (and unique within each record)* | 184,228 | 59,976 | 101,231 | 169,417 | 69,244 | 69,957 | 99,660 | 60,896 | 68,625 | 192,412 | 117,138 | 68,061 | 24,560 | 12,068 | 10,774 | 17,972 | 1,302 | 9,020 |
| *Misassigned ISBNs* | 84,817 | | | 60,272 | | | 56,121 | | | 56,217 | | | 8,909 | | | 7,285 | | |
| *Final PID list size* | 260,618 | | | 248,346 | | | 173,060 | | | 321,394 | | | 38,493 | | | 21,009 | | |

## Results

By processing the IRIS dumps of the six involved institutions, we have obtained more than 1.6M records – 402,505 records in UNIBO IRIS, 381,525 in UNIMI IRIS, 326,137 in UNITO IRIS, 416,547 in UNIPD IRIS, 62,638 in UPO IRIS, and 39,979 SNS IRIS. A detailed distribution of these records per year (i.e. record creation in IRIS) is shown in Figure 6. The increase observed after 2000 reflects a policy introduced in Italy to run a nationwide research assessment exercise for universities and other research institutions called *Valutazione della Qualità della Ricerca* (VQR), i.e., *Research Quality Evaluation* (https://www.anvur.it/en/research/evaluation-research-quality), which was conducted for the very first time in 2011 on all publications by Italian research institutions from 2004 to 2010. The relatively low number of records for 2025 and 2026 is instead explained by the fact that the IRIS dumps were provided to us by the institutions involved over different periods, from May to October. Thus, they do not give a complete snapshot of the research outcomes produced by universities by the end of 2025. Therefore, to avoid potential data loss, all analyses presented here consider all bibliographic records listed in IRIS installations published by 2024 (inclusive).

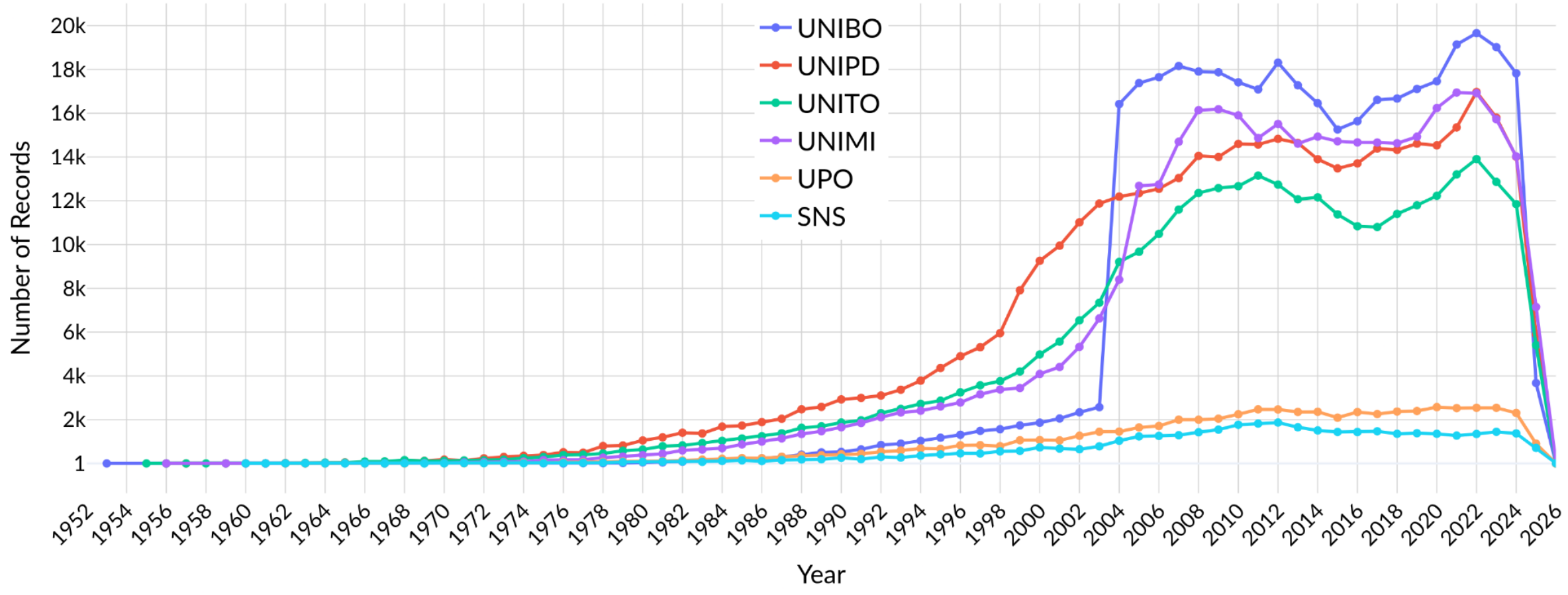

**Figure 6.** Distribution of the publications per year in the IRIS installations of six different Italian universities.

**Table 4.** IRIS records matched in OpenCitations Meta across the six institutions. Percentages are computed with respect to the total number of records in each IRIS data dump for which the publication date is equal to or less than 2024.

| UNIBO | UNIPD | UNITO | UNIMI | UPO | SNS |
|---|---|---|---|---|---|
| 165,500 (42.7%) | 161,843 (39.8%) | 100,915 (31.8%) | 176,262 (48.1%) | 23,235 (38.0%) | 15,988 (41.6%) |

As described in Section “Methodology”, from these IRIS records, we extracted all publications with *valid* (i.e., syntactically correct) persistent identifiers (PIDs) compliant with OpenCitations Meta: DOI, PMID, or ISBN. A summary of PID extraction and validation outcomes for each IRIS data dump is reported in Table 3. Instead, Table 4 reports the number and proportion of IRIS records successfully matched within OpenCitations Meta for each university involved in the study (RQ1). Overall coverage ranges from 31.8% (UNITO) to 48.1% (UNIMI), with an average of 40.33% (median: 40.70%).

**Table 5.** Citations involving IRIS records (as citing/cited entities or both) in the OpenCitations Index across the six institutions. The number in parentheses indicates the average number of outgoing/incoming citations for each IRIS record mapped in OpenCitations Meta.

| | UNIBO | UNIPD | UNITO | UNIMI | UPO | SNS |
|---|---|---|---|---|---|---|
| *IRIS records as citing entity* | 5,263,041 (31.78) | 5,305,202 (32.78) | 3,604,946 (35.72) | 5,922,327 (33.56) | 823,215 (35.43) | 460,882 (28.87) |
| *IRIS records as cited entity* | 5,192,299 (31.35) | 5,416,533 (33.47) | 3,640,514 (36.07) | 7,042,335 (39.95) | 829,897 (35.72) | 523,012 (32.71) |
| *Citations between IRIS records* | 462,695 | 474,730 | 268,361 | 494,495 | 57,094 | 52,540 |

Table 5 summarises the number of citations in the OpenCitations Index that involve IRIS records (RQ2). The numbers vary proportionally with the total number of mapped entities retrieved (Table 4). Indeed, the average number of entities referenced by the IRIS records mapped in OpenCitations Meta is 33.02 (median: 33.17), whereas the average number of citations received per entity is 34.88 (median: 34.60).

Finally, Figure 7 shows the distribution of all IRIS records that have not been found in OpenCitations Meta, either because no IRIS entity PIDs were found or because the IRIS entity lacked all relevant PIDs used in the study (RQ3). As shown in the data, the primary publication types missing are journal articles, book chapters, and proceedings papers, even though they have high coverage across all IRIS installations and are the most represented in absolute numbers. Instead, the publication types with coverage of less than 10% in OpenCitations Meta across at least five IRIS installations are *book translation*, *concordance*, *critical edition*, *database*, *editorship*, *monograph or scientific treatise*, and *scientific commentary*, all of which are different kinds of books except *database*. This shows that OpenCitations Meta currently lacks a significant amount of literature from the involved universities that has been traditionally published in the Social Sciences and Humanities domain.

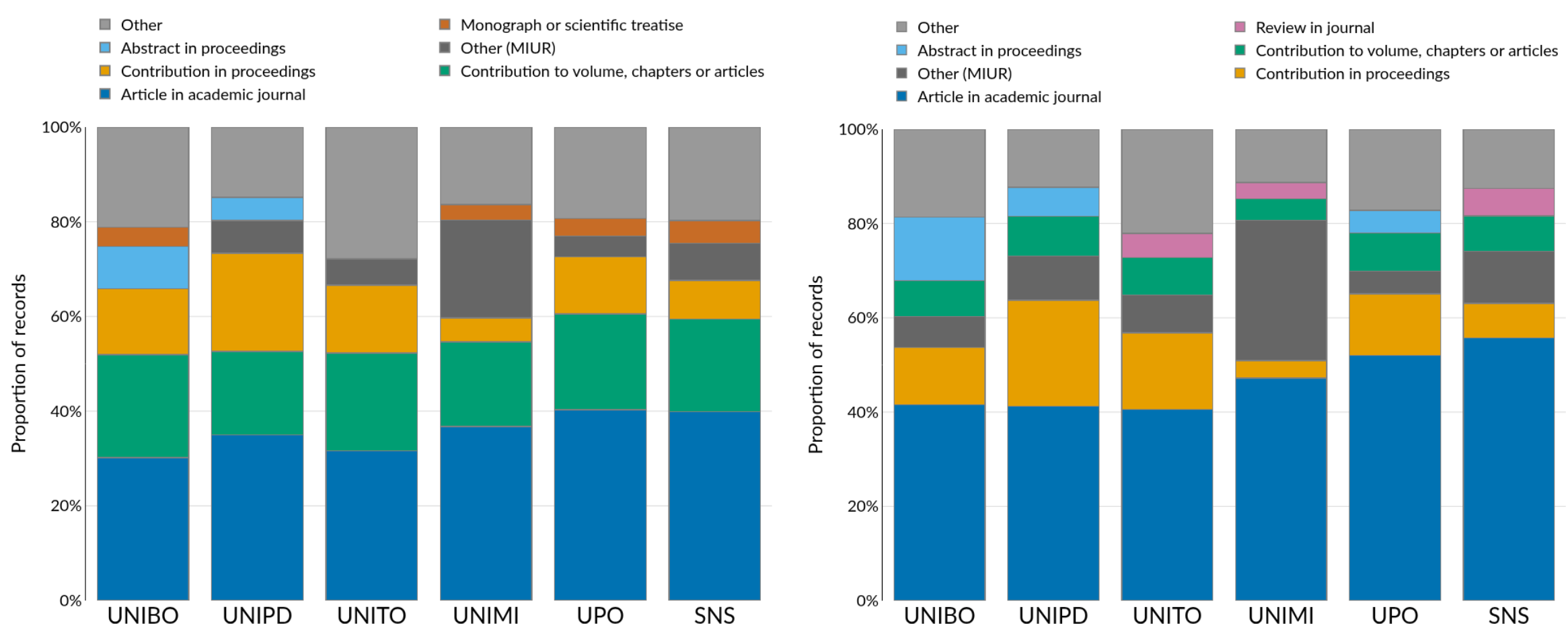

**Figure 7.** Distribution of MIUR publication types among (left) IRIS records having PIDs that have not been found in OpenCitations Meta, and (right) IRIS records without PIDs that were not mapped to OpenCitations Meta. Only the five most frequent publication types are shown explicitly; the remaining types are aggregated into the "Other" category for visualisation purposes. It is worth clarifying that the "Other" used in the graphs above should not be confused with the "Other (MIUR)" category, which is the residual category we used when either IRIS installations refer explicitly to it, or we did not find any correspondence between the IRIS publication type and the MIUR publication-type classification.

# Discussions

The results obtained indicate a homogeneous situation across the participating institutions. Indeed, considering also the sources OpenCitations uses to retrieve and ingest the data it makes available, the best coverage concerns publication types that are traditionally widely distributed across several disciplinary fields. The best covered publication type in OpenCitations Meta is *proceedings paper* (UNIBO: 84.19%, UNIMI: 85.38%, UNIPD: 83.49%, UNITO: 74.04%, UPO: 78.31%, SNS: 87.59%; average: 82.16%, median: 83.84%), followed by *book chapters* (UNIBO: 69.47%, UNIMI: 68.07%, UNIPD: 74.44%, UNITO: 69.55%, UPO: 75.81%, SNS: 65.28%; average: 70.44%, median: 69.51%) and *journal articles* (UNIBO: 67.39%, UNIMI: 57.93%, UNIPD: 67.13%, UNITO: 62.83%, UPO: 64.29%, SNS: 85.25%; average: 67.47%, median: 65.71%). Journal articles are, overall, the most represented category across the various IRIS installations, despite having limited coverage. Indeed, there are specific articles and journals included in IRIS that OpenCitations Meta does not cover.

From a previous study conducted on UNIBO IRIS data with a similar timeframe (Andreose et al., 2026a) that was more inclusive (it considered publications and citations by 1 April 2025) than the work presented here, we can extract the coverage of IRIS publication entities in Scopus and Web of Science, which were 144,940 (36%) and 129,823 (32.25%), respectively. These values are smaller than those shown in Table 4, which is 165,500 (42.7%). Even if such information comes from only one of the institutions involved, given the homogeneity of Italian institutions, this finding suggests that OpenCitations may already have better coverage of IRIS entities than the two proprietary services.

Along the same lines, comparing the results in Table 5 with those reported by Andreose et al. (2026a) for Scopus and Web of Science for UNIBO shows a promising picture. In fact, Scopus reported 5,225,193 citations for UNIBO entities, with an average of 36.05 citations per entity. In contrast, Web of Science reported 4,505,715 citations (average of 34.71 per entity), which are very close to those shown in Table 5 (5,192,299 citations, with an average of 31.35 per entity). Even though these results show fewer citations tracked for the publication entities included in OpenCitations Meta than in Scopus and Web of Science, the number of citations retrieved is comparable. Given that Italian institutions show similar behaviour in publication distributions and citation counts – indeed, UNIBO had the worst average citation per entity value – this finding may suggest the potential adoption of OpenCitations as an alternative to proprietary citation-count services.

These results also open up interesting avenues for future work from the OpenCitations perspective. One potential strategy to increase the coverage of ingested resources is the crowdsourcing approach initially proposed by Heibi et al. (2019). This method is currently under active consideration by the OpenCitations team, which is developing an automated workflow to support it. The approach encourages researchers and institutions to contribute bibliographic metadata, which can then be processed through a structured pipeline that includes quality assessment and validation of the submitted data (Heibi et al., 2025). Technically, this approach would enhance OpenCitations Meta and OpenCitations Index by incorporating additional resources — bibliographic records and citation data — that were not previously covered by the traditional sources used by OpenCitations (e.g. Crossref).

**Interpreting the ‘Other (MIUR)’ publication category**

A relevant aspect emerging from the analysis concerns the interpretation of the “Other (MIUR)” category used in the classification of publication types. This category includes heterogeneous records originating from two distinct sources.

First, IRIS itself allows institutions to define local publication types, including generic or residual categories (e.g. “other”), reflecting the system's flexibility in accommodating institutional needs. Second, during alignment with the national classification defined by the MIUR, we mapped all IRIS publication types that lacked a clear correspondence to the residual MIUR category “Other (MIUR)” to ensure comparability across institutions. As a consequence, the resulting “Other (MIUR)” category aggregates records that are heterogeneous by nature, including (i) locally defined publication types with no direct equivalent in the MIUR taxonomy, and (ii) cases in which a one-to-one mapping was not possible due to ambiguity. This aggregation may therefore mask structurally different types of scholarly outputs.

This issue is particularly evident in UNIMI data, which shows a comparatively high proportion of records classified as “Other (MIUR)”. In the UNIMI IRIS implementation, the local customisation of publication types includes a complex hierarchy of categories and subcategories that partially overlap and do not map cleanly to the MIUR classification. In several cases, individual IRIS types could correspond to multiple MIUR categories, while other subcategories had no direct equivalent. This made it difficult to assign records unambiguously, leading to a larger proportion of entries being grouped under the residual category.

These findings highlight a broader limitation of CRIS systems such as IRIS: while local customisation enables institutions to represent their research outputs better, it may hinder interoperability and comparability when data are aggregated across institutions or aligned with standard taxonomies. A more systematic harmonisation of publication-type classifications, or the adoption of shared controlled vocabularies, would improve the interpretability of cross-institutional analyses such as those presented in this study.

**Limitations of the approach**

The context in which the providers of open research information operate is closely tied to the Open Science ecosystem envisioned by the scholarly community and supported by major national and international institutions such as UNESCO and CoARA. It is clear among participants in this discourse that it is impossible to have a single, open scholarly infrastructure (POSI Adopters, 2025) capable of hosting all available research information for any use. That is why the scholarly community is working on a decentralised approach in which several open infrastructures, distributed across the globe and handling partially overlapping data to guarantee their reliability, should be seen as a single meta-entity, e.g., a *meta-infrastructure*, able to provide all the needed open research information. Thus, the analysis we have described in this article can have additional evidence coming from other infrastructures in addition to OpenCitations that, speculatively speaking, would provide even more positive results to those addressed here, which we believe are already a good signal to justify the feasibility of switching from a closed system to open infrastructures.

In addition, the possibility of a fair comparison with closed, proprietary systems is not easy to address. Indeed, in (Andreose et al., 2026a), we had the opportunity to compare, within a particular institution (UNIBO), quantitative counts of publication coverage and citations with those in Scopus and Web of Science. Data from such proprietary services were retrieved by asking the competent university office to gather them via their respective APIs, specifying the Scopus ID and the Web of Science ID for each publication included in UNIBO IRIS, since these identifiers are stored within the IRIS installation of the University of Bologna. Thus, in the present article, we have used the conclusions drawn by Andreose et al. (2026a) as a baseline for comparison with the other Italian universities involved in the study. We are aware that, ideally, a fairer comparison can be made only by retrieving the specific information from Scopus and Web of Science for all the universities involved, even though this was not easy to set up systematically. This is primarily because the data we could access via Scopus and Web of Science APIs is regulated by national agreements (signed by all Italian universities), which allowed us to access only limited information, which excludes, for instance, information about the entities citing those in the IRIS installations, thus not preventing us on having more appropriate comparisons in terms of citation coverage. In addition, as a consequence of the same signed agreement at the national level, we are not legally entitled to publish any raw data from Scopus or Web of Science, thereby preventing the full reproducibility of any conclusions we may derive from them.

**Software for analysing open research information**

Developing and providing open software to support universities and other research-performing organisations, and funders, in autonomously evaluating the coverage and quality of providers of open research information is key in this context. Indeed, our goal was not only to run an exploratory study within the Italian university system, but also to provide an easy-to-install software (Zilli et al., 2025) that could be used independently by any Italian university to repeat the analysis and experimentation in the future with their own IRIS data.

Offering tools and instruments to the community is one of the most valuable advantages that initiatives such as the Barcelona Declaration aim to establish, enabling actors to make informed choices. Fortunately, in recent years, we have seen research groups publish several tools to support these purposes. For instance, Nikolić et al. (2024) have recently proposed an open-source tool for enabling the automatic preprocessing and deduplication effort of records based on similarity and other user-configurable strategies that can be used, in principle, to mash up data coming from different sources – in their study, for instance, they showed an application of records from Scopus and Web of Science. Another recent, complementary tool for gathering information from existing sources was *pyBiblioNet* (Lai et al., 2025). This Python wrapper interacts with the OpenAlex API to gather data and provides implemented pipelines for preprocessing, visualising, and analysing bibliometric data from OpenAlex (https://openalex.org/), another open scholarly infrastructure providing bibliographic information about the global scholarly production. Investments in these kinds of tools to simplify access to and analysis of open research information will become crucial to supporting the transition from closed systems to open sources.

## Conclusions

The analysis introduced in the article provides an initial overview of the coverage of Italian scholarly publications in OpenCitations. This Open Science infrastructure provides open research information, in particular, bibliographic metadata and citation data. Our findings suggest the premises are already in place to replace proprietary services with open research information providers. While here we focused on the primary insights from the analysis, the data available on Zenodo (Andreose et al., 2026b) also includes HTML-based reports for all the universities involved, which include additional tables and visuals.

Indeed, we have obtained promising results by using data from only one Open Science infrastructure in the analysis. However, OpenCitations is in good company within the scholarly systems, which includes, to mention a few, OpenAIRE (https://www.openaire.eu/), GoTriple (https://www.gotriple.eu/), OpenAlex, Crossref (https://www.crossref.org/), DataCite (https://datacite.org/), Directory of Open Access Journals (DOAJ, https://doaj.org/), Directory of Open Access Books (https://www.doabooks.org/), OAPEN (https://www.oapen.org/), and Thoth (https://thoth.pub/).

Of course, integrating data from different infrastructures is not necessarily easy, even when supported by existing software, particularly when each infrastructure uses its own mechanisms to access its data and exposes them using non-standard data models and formats. Since we have started building the federation of Open Science infrastructures in the past years, e.g. in the context of the European Open Science Cloud (EOSC, https://eosc.eu/), there is an urgent need to enable interoperability between infrastructures. The scholarly community has begun to push initiatives to enable easier data mashups across different sources. For instance, recently, the *Scientific Knowledge Graph – Interoperability Framework* (SKG-IF, https://skg-if.github.io/) (Mannocci et al., 2025) has provided specifications (i.e., data model, formats, REST API) to address technical and semantic interoperability among providers of open research information. Instead, the Collaborative Metadata initiative (COMET, https://www.cometadata.org/) recently launched to build a community-based approach to metadata enrichment, aiming at improving the quality and coverage of the open metadata made available by existing open infrastructures.

Finally, as future perspectives, we want to extend our analysis in different directions. First, one aspect to investigate in depth is measuring the coverage of IRIS publications and citations in Scopus and Web of Science. While in this article we have based our speculation in this direction on a prior analysis conducted at the University of Bologna (Andreose et al., 2026a), quantitative data from all other universities in this respect can provide more quantitative clues. In addition, we will continue to encourage other Italian universities, in particular those located in the Central and Southern parts of Italy, to have a more balanced distribution of the data sources, to participate in this analysis by requesting their IRIS data dumps, aiming to create a more complete snapshot of the current Italian scholarly system. This is facilitated by the development of the software used in the study (Zilli et al., 2025), which can also run similar analyses on data exposed in other non-Italian CRIS, provided the input is prepared in a format compliant with the CSV format we adopted.

# Appendix 1

This appendix lists the seven SQL queries provided to all participating universities to extract relevant information from their IRIS installations for our analysis.

**Query 1: actors' metadata**

```
select ITEM_ID,RM_PERSON_ID,PID,ORCID,FIRST_NAME,LAST_NAME,PLACE

from ODS_L1_IR_ITEM_CON_PERSON
```

**Query 2: author lists**

```
select
ITEM_ID,DES_ALLPEOPLE,DES_ALLPEOPLEORIGINAL,DES_NUMBEROFAUTHORS,DES_
NUMBEROFAUTHORS_INT

from ODS_L1_IR_ITEM_DESCRIPTION
```

**Query 3: publication's identifiers**

```
select
ITEM_ID,IDE_DOI,IDE_EISBN,IDE_ISBN,IDE_ISBN_1,IDE_ISBN_2,IDE_ISBN_3,
IDE_ISMN,IDE_OTHER,IDE_PATENTNO,IDE_PATENTNOGR,IDE_PATENTNOPB,IDE_PM
ID,IDE_SOURCE,IDE_UGOV,IDE_URL,IDE_URL_1,IDE_URL_2,IDE_URL_3,IDE_CIT
ATION

from ODS_L1_IR_ITEM_IDENTIFIER
```

**Query 4: publication's language**

```
select *

from ODS_L1_IR_ITEM_LANGUAGE
```

**Query 5: publication's basic bibliographic metadata**

```
select
ITEM_ID,DATE_ISSUED_YEAR,TITLE,OWNING_COLLECTION,OWNING_COLLECTION_D
ES

from ODS_L1_IR_ITEM_MASTER_ALL
```

**Query 6: publisher's metadata**

```
select ITEM_ID,PUB_NAME,PUB_PLACE,PUB_COUNTRY,PUB_COUNTRY_I18N

from ODS_L1_IR_ITEM_PUBLISHER
```

**Query 7: publication's context metadata**

```
select
ITEM_ID,REL_ALLAUTHORS,REL_ALLAUTHORSORIG,REL_ALLEDITORS,REL_ARTICLE
,REL_CONFERENCEDATE,REL_CONFERENCENAME,REL_CONFERENCENUMBER,REL_CONF
ERENCEPLACE,REL_CONFERENCETARGETAUDIENCE,REL_CORPORATE,REL_EDITION,R
EL_FIRSTPAGE,REL_FORMAT,REL_FUND,REL_ISPARTOFBOOK,REL_ISPARTOFJOURNA
L,REL_ISPARTOFJOURNAL_CRIS_ID,REL_ISPARTOFJOURNAL_ANCE,REL_ISPARTOFJ
OURNAL_IS_SCIENT,REL_ISPARTOFJOURNAL_SCIEN_AREA,REL_ISPARTOFJOURNAL_
DOAJ,REL_ISPARTOFJOURNAL_PUBLISHER,REL_ISPARTOFSERIE,REL_ISPARTOFSER
IE_CRIS_ID,REL_ISPARTOFSERIE_ANCE,REL_ISPARTOFSERIE_IS_SCIENT,REL_IS
PARTOFSERIE_SCIEN_AREA,REL_ISPARTOFSERIE_PUBLISHER,REL_ISSN,REL_ISSN
_IN_ERIH_PLUS,REL_ISSUE,REL_LASTPAGE,REL_MEDIUM,REL_NUMBEROFPAGES,RE
L_PROJECT,REL_PROJECT_AUTHORITY,REL_PROJECT_CRIS_ID,REL_VOLUME,REL_P
ROJECT_DESCRIPTION,REL_PROJECT_COUNT,REL_PROJECT_FUNDER_NAME,REL_PRO
JECT_FUNDING_STREAM,REL_PROJECT_TITLE,REL_PROJECT_NUMBER

from ODS_L1_IR_ITEM_RELATION
```

## Data availability statement

The datasets used and analysed during the current study are freely available on Figshare (https://figshare.com/) and Zenodo (https://zenodo.org). In particular:

**OpenCitations Meta** – Published on June 9, 2025 (10.5281/zenodo.15625651). The dataset comprises 124,526,660 bibliographic entities, 376,295,095 authors, 2,765,927 editors, 1,019,563 publication venues, and 103,928,927 publishers. The compressed data totals 12 GB (49 GB when uncompressed) and is distributed across 38,602 CSV files.

**OpenCitations Index** – Published on July 15, 2025 (10.6084/m9.figshare.24356626.v6). The dataset comprises 2,216,426,689 citations. The compressed data totals 38.8 GB, while the unzipped CSV files total 242 GB.

**IRIS installations** – All data gathered from the six IRIS installations can be downloaded freely from the related websites (UNIBO: https://cris.unibo.it/, UNIMI: https://air.unimi.it/, UNIPD: https://www.research.unipd.it/, UNITO: https://iris.unitn.it/, UPO: https://iris.uniupo.it/, SNS: https://ricerca.sns.it/).

**Results** – Published on January 9, 2026 (https://doi.org/10.5281/zenodo.18202530). Contains all the data produced as a consequence of the analysis described in this work.

**Software** – Published on December 23, 2025 (https://doi.org/10.5281/zenodo.18040113). All code used for data processing and analysis for the current study is openly available on GitHub (https://github.com/opencitations/iris-oc-mapper) and archived on Zenodo.

## Authors' contribution statements

Erica Andreose: Conceptualization, Investigation, Methodology, Validation, Visualization, Writing – original draft, Writing – review & editing

Ivan Heibi: Resources, Methodology, Supervision, Validation, Writing – original draft, Writing – review & editing

Silvio Peroni: Conceptualization, Funding acquisition, Methodology, Resources, Supervision, Validation, Writing – original draft, Writing – review & editing

Leonardo Zilli: Conceptualization, Data curation, Investigation, Methodology, Software, Validation, Writing – original draft, Writing – review & editing

## Acknowledgements

This work has been partially funded by the European Union's Horizon Europe framework programme under Grant Agreements No 101095129 (GraspOS Project) and No 101187940 (LUMEN Project).

The authors want to thank all the people from the universities involved in the analysis who facilitated and/or provided us with the data to run the analysis: Mauro Apostolico (UNIPD), Cristiana Bettella (UNIPD), Stefano Bolelli Gallevi (UNIMI), Paola Galimberti (UNIMI), Elena Giachino (UNIBO), Giorgio Longo (UNITO), Alberto Massarotti (UPO), Ennio Misuraca (UNIBO), Patrizia Parisi (UNITO), and Donatella Tamagno (SNS).

## Conflict of interest

SP is the University of Bologna's representative for the Barcelona Declaration, Director of OpenCitations, and Co-Chair of the RDA Working Group on Scientific Knowledge Graph – Interoperability Framework. IH is the Chief Technology Officer of OpenCitations. OpenCitations is one of the open infrastructures that formally supports the Declaration.